\documentclass[conference]{IEEEtran}
\usepackage{amsmath, amssymb,multicol, graphicx,gensymb, flushend}

\IEEEoverridecommandlockouts

\begin{document}

\title{Catalytic Nanoparticles: An Introduction}

\author{\IEEEauthorblockN{Jairo Rondón\IEEEauthorrefmark{1},
Angel Gonzalez-Lizardo\IEEEauthorrefmark{2}, Claudio Lugo\IEEEauthorrefmark{3}}
\IEEEauthorblockA{\IEEEauthorrefmark{1}Biomedical \& Chemical Engineering Departments,\\
Polytechnic University of Puerto Rico, San Juan, Puerto Rico, USA\\jrondon@pupr.edu}
\IEEEauthorblockA{\IEEEauthorrefmark{2}Department of Electrical and Computer Engineering and Computer Science,\\
Polytechnic University of Puerto Rico, San Juan, Puerto Rico, USA\\agonzalez@pupr.edu}
\IEEEauthorblockA{\IEEEauthorrefmark{3}Laboratorio de Cinética y Catálisis, Universidad de Los Andes, Mérida, Venezuela\\claudiolugo@ula.ve}
}

\maketitle

\begin{abstract}

This study explores the transformative potential of nanocatalysts, emphasizing their pivotal role in catalysis and material science. Key synthesis techniques, including chemical reduction and hybrid methods, are highlighted for their ability to control particle size and enhance stability. Applications in environmental remediation, fuel quality improvement, and renewable energy showcase the broad impact of nanocatalysts. 
Despite challenges in scalability and stabilization, advancements in bimetallic configurations and electro-steric approaches demonstrate significant progress. This research underscores nanocatalysts' promise for sustainable industrial processes and global challenges.

\end{abstract}

\section{Introduction}

In the last decade, nanoscience and nanotechnology have emerged as transformative fields globally in research and development. Advances in manipulating materials at the nanoscale have fundamentally changed how materials, devices, and systems are designed and understood. Nanotechnology is based on the use of materials and systems at the atomic scale, specifically at the nanometer level (one nanometer is equal to one billionth of a meter) \cite{r1}. A clear example of its potential is observed in nanocatalysis, where chemical reactions are precisely controlled by manipulating the size, composition, and morphology of reactive centers. This subfield demonstrates significant impacts on reaction kinetics, industrial processes, and energy applications \cite{r2}. The purpose of this review is to explore the potential of nanoparticles, particularly their applications in catalysis. Transition metal nanoparticles exhibit exceptional catalytic activity in organic reactions and advanced industrial processes. Understanding these materials can significantly enhance energy efficiency and sustainability \cite{r3}

\section{Methodology}
For this research, a qualitative-documentary methodology was employed, based on:
\begin{enumerate}
\item Search and data compilation: Textbooks and databases such as SciELO, RedAlyC, and Google Scholar were utilized. Key search terms included ``Nanocatalysis'', ``Catalytic nanoparticle'', and ``Nanomaterial''.
\item Selection and refinement of information: A comprehensive exploration was conducted for the period 1970 - 2023. Mendeley \cite{r4} was used as a bibliographic management tool to organize data based on their relevance to the four fundamental frameworks of this study: conceptualization, synthesis methods, stabilization techniques, and applications in nanocatalysis.
\item Selection of subtopics: The refined information facilitated structuring the research and clarifying the subtopics chosen for the study.
\item Analysis of results: The data were critically analyzed, leading to comprehensive conclusions.
\end{enumerate}

\section{Nanoscience and Nanotechnology: An Integrated Overview}
Nanoscience is the field dedicated to the study of phenomena and properties of materials at the nanoscale, a dimensional range where unique interactions emerge. This scale, typically ranging from 1 to 100 nanometers, defies traditional categorizations of physics, chemistry, and biology, as it embodies a convergence of these disciplines. At this level, complex interactions dominate, providing insights into behaviors that are not observed in macroscopic systems \cite{r}.
Breakthroughs in experimental techniques such as scanning tunneling microscopy (STM) and atomic force microscopy (AFM) have enabled researchers to observe and manipulate individual atoms and molecules \cite{r6}. These technological advancements have significantly enhanced our ability to design and explore materials at the atomic scale. Beyond expanding fundamental scientific knowledge, nanoscience has catalyzed innovations in electronics, energy, and medicine. For instance, the development of quantum dots and metallic nanoparticles has transformed fields such as optoelectronics and chemical catalysis, showcasing how nanoscale control can yield unique properties like high reactivity, selectivity, and thermal stability \cite{r7},\cite{r8},\cite{r9}.
Nanotechnology, closely tied to nanoscience, involves the controlled manipulation and fabrication of materials, devices, and systems at the atomic and molecular levels. Its historical roots can be traced to the pioneering experiments of Michael Faraday in the 19th century, who explored the behavior of metallic nanoparticles, and the theoretical work of Gustav Mie, who explained the optical properties of glass embedded with nanoparticles. The concept of nanotechnology gained prominence in 1959 when Nobel laureate Richard Feynman envisioned constructing materials and devices by rearranging atoms and molecules with precision \cite{r10}.
The field advanced significantly in the 1980s through Eric Drexler’s introduction of ``molecular nanotechnolog''. This concept proposed building intricate structures atom by atom using nanoscale machines \cite{r11}. Drexler’s vision inspired the development of transformative technologies, such as carbon nanotubes, nanocapsules for targeted drug delivery, and nanowires for electronic applications. Today, nanotechnology is recognized as a profoundly interdisciplinary domain, integrating principles from chemistry, physics, biology, and engineering to tackle global challenges \cite{r12}.
The transformative potential of nanotechnology lies in the substantial changes in the physical and chemical properties of materials when they are scaled down to the nanoscale. This phenomenon, known as the quantum effect \cite{r13}, gives rise to novel behaviors absent in bulk materials. Key properties include:
\begin{itemize}
\item Electrical conductivity: Significantly enhanced in materials like carbon nanotubes.
\item Chemical reactivity: Increased, enabling advances in both heterogeneous and homogeneous catalysis.
\item Mechanical strength: Improved in polymer composites reinforced with nanoparticles.
\item Optical properties: Exemplified by quantum dots and noble metal nanoparticles.
\end {itemize}
These properties have driven transformative applications \cite{r14},\cite{r15},\cite{r16}, such as:
Controlled drug delivery systems.
Lightweight, durable materials for aerospace and automotive industries.
High-efficiency, miniaturized electronic devices.
Although the term ``nanotechnology'' is relatively recent, the application of nanomaterials has a long history. Roman artisans in the 4th century and medieval craftsmen employed metallic nanoparticles to produce vibrant colors in glassware and stained-glass windows.
In the 19th century, Faraday synthesized metallic nanoparticles, and Ostwald proposed theoretical models emphasizing the importance of surface atoms in nanoscale materials. The modern era of nanotechnology began with Feynman’s groundbreaking lecture, There’s Plenty of Room at the Bottom, which established the foundational vision for manipulating matter at the atomic level \cite{r17}. Subsequent innovations in microscopy, including STM and AFM during the 1980s, provided researchers with the tools to observe and manipulate materials at this scale, inaugurating a new chapter in nanoscience and nanotechnology.
This continuous evolution positions nanotechnology as a pivotal force in addressing 21st-century challenges across science, technology, and industry \cite{r18}.
\section{Catalytic Nanoparticles}

Nanomaterials represent an emerging class of materials that have revolutionized the field of materials science due to their unique properties and their impact on various industrial and technological applications. These materials can be metallic, semiconducting, polymeric, ceramic, or combinations thereof. The reduction in the size of these materials induces significant effects on their mechanical, thermal, optical, and electrical properties, setting them apart from their macroscopic counterparts \cite{r19}, \cite{r20}, \cite{r21}.
One of the fundamental concepts in materials science when studying nanomaterials is the impact of size reduction on their surface energy. As the size of a particle decreases to the nanometric scale, the number of atoms on the surface increases exponentially compared to those in the interior. This results in higher surface energy due to the low coordination of surface atoms, which imparts unique physical and chemical properties to the material, such as enhanced chemical reactivity, wear resistance, and improved thermal stability \cite{r22}, \cite{r23}.
Nanoparticles, defined as isolated particles with sizes ranging from 1 to 50 nanometers, occupy an intermediate space between atoms and molecules and bulk materials. These nanoparticles exhibit properties such as a high surface-to-volume ratio, which enhances their efficiency in catalytic applications. For example, in heterogeneous catalysis, nanoparticles of noble metals like gold, platinum, or palladium can increase the speed and selectivity of chemical reactions. This is because their high density of active sites facilitates specific interactions at the molecular level, reducing energy barriers and improving the overall efficiency of the process \cite{r24}, \cite{r25}.
Another key concept is that of clusters or aggregates, which consist of groups of atoms ranging in size from $2$ to $10 \times 10^{4}$ atoms. These clusters exhibit unusual structures and distinctive properties that differ from macroscopic crystalline solids, making them ideal for specific applications such as advanced catalysts. In particular, metallic clusters have proven essential in homogeneous and heterogeneous catalysis, where their nanometric structures provide a larger active surface for reactions such as oxidation-reduction, hydrogenation, and catalytic cracking \cite{r25}, \cite{r26}.
In the context of nanocomposites, nanomaterials are also used as reinforcements to improve the mechanical, thermal, and catalytic properties of base materials. Nanocomposites based on metal oxides, such as titanium oxide and cerium oxide, are widely used as catalytic supports in energy conversion reactions and industrial processes, due to their chemical stability and high capacity to disperse active nanoparticles \cite{r28}, \cite{r29}. 
Nanomaterials with catalytic properties also excel in optical and electronic applications. For example, semiconductor quantum dots can be used in photocatalysis for pollutant decomposition processes or hydrogen generation through water splitting. Additionally, metallic nanoparticles with plasmonic properties are employed to enhance light harvesting and the efficiency of catalytic solar devices \cite{r30}.
Ultimately, nanomaterials and nanoparticles with catalytic properties represent a key tool in materials science for developing innovative and sustainable solutions in sectors such as energy, the chemical industry, and electrical, mechanical, and environmental engineering. Understanding the principles governing their properties and catalytic behavior enables the optimization of their design and application across a wide range of technologies, solidifying them as pillars of modern materials research.

\subsection{Classification and Properties}

Nanomaterials are classified based on the number of dimensions confined within the nanometric range, which defines their physicochemical behavior and technological applications. Zero-dimensional (0-D) materials, such as metallic, semiconducting, and metal oxide nanoparticles, have all their dimensions confined to the nanometric range, granting them unique properties such as electronic quantization effects and high surface activity. On the other hand, one-dimensional (1-D) materials, such as carbon nanotubes and metallic nanowires, are notable for their anisotropic properties, making them ideal for applications in advanced electronic devices and sensors. Two-dimensional (2-D) materials, such as graphene, thin films, and nanometric membranes, combine a high surface-to-volume ratio with exceptional mechanical and electronic properties, making them essential in the manufacture of flexible devices and molecular separation technologies. Finally, three-dimensional (3-D) materials, such as self-assembled networks and three-dimensional porous structures, offer opportunities in energy storage, catalysis, and bioengineering due to their structural and functional complexity  \cite{r31}.
Moreover, Gleitter proposes a structural classification that includes layered systems, consolidated systems, and nanoparticles supported on solid or liquid matrices, broadening the possibilities for manipulation in specific applications. These categories allow for rational design to optimize key properties such as thermal stability, mechanical resistance, and catalytic activity \cite{r32}, \cite{r33}. In terms of catalytic properties, nanoparticles exhibit a high surface-to-volume ratio, exposing a significant proportion of atoms on their surface, which enhances reactant adsorption and catalytic interactions. This, along with quantum effects generated by electron confinement, influences bonding energies and the density of electronic states, altering chemical reactivity. High catalytic activity is attributed to the abundance of active sites at grain boundaries, surface defects, and specific facets that promote preferential reaction pathways. Additionally, their nanometric design allows for controlled selectivity, facilitating the conversion of reactants into desired products while minimizing unwanted by-products, making them ideal for sustainable processes.
These catalytic nanoparticles can also be classified by their origin. Naturally occurring nanoparticles include biological systems such as proteins and viruses, or those formed through geological and environmental processes, such as volcanic dust. In contrast, human-made nanoparticles can be generated unintentionally during industrial processes (e.g., pyrolysis, combustion) or deliberately through advanced nanotechnology techniques such as chemical vapor deposition (CVD) or sol-gel synthesis. The latter have proven essential in petrochemical industry applications, reducing pollutant emissions, and developing materials with customizable optical and catalytic properties.
This combination of classification and properties makes catalytic nanoparticles fundamental tools for addressing technological challenges in heterogeneous catalysis, energy conversion, and biomedical applications. Their versatility and ability to be custom-designed ensure their relevance in advancing modern science and engineering \cite{r34}.

\subsection{Size and Structure Dependence}
The physicochemical properties of materials vary significantly depending on the dimensional scale at which they are analyzed. At macroscopic scales, material properties are considered averaged, such as density, elastic modulus, resistivity, magnetization, and dielectric constant. However, when the scale of study is reduced to the micrometric and nanometric range, these properties can undergo drastic changes, particularly in mechanical, ferroelectric, and ferromagnetic aspects. In the nanometric range, between 1 and 100 nm, nanoparticles exhibit unique behavior due to phenomena such as relative surface increase and quantum confinement, which requires understanding microscopic and mesoscopic properties as well to contextualize the observed phenomena \cite{r1}, \cite{r2}, \cite{r8}.
\begin{itemize}
\item Internal Energy: According to Poltorak and Van Hardeveld, metal nanoparticles smaller than 50 Å exhibit significantly different properties compared to their massive counterpart. This is due to the increased proportion of surface atoms to total volume, resulting in lower coordination of surface atoms and, therefore, higher surface energy. This characteristic makes them highly reactive and gives them unique properties for catalytic applications \cite{r35}, \cite{r36}, \cite{r37}.
\item Reactivity: The relationship between nanoparticle size and reactivity is neither linear nor unbounded \cite{r38}. Studies such as those conducted by Liu et al. on palladium nanoparticles have shown that although smaller nanoparticles have higher surface energy, their reactivity towards certain reactions, such as hydrogen dissociation, can be decreased \cite{r39}. This is because certain reactions require specific structures on the catalytic surface, and reducing the size can alter the geometry and structure of the nanoparticles, affecting their effectiveness. Reactivity therefore depends on a complex interplay between size, surface structure, and reaction type \cite{r40}, \cite{r44}.
\item Metal-Support Interaction: In supported nanoparticles, such as niquel-cobalto in MgO \cite{r9}, alumina, or NaY zeolite, the interaction between the metal and the support plays a critical role \cite{r41}. Although small nanoparticles are typically more reactive, they are also more likely to interact with the support, which stabilizes the nanoparticles and reduces their catalytic activity. Boudart argues that there is an optimal supported nanoparticle size to maximize catalytic activity. This balance highlights the importance of controlling not only the size of the nanoparticles, but also the characteristics of the support, to optimize their performance in specific applications \cite{r42}, \cite{r43}.
\end{itemize}

The dependence of physicochemical properties on nanoparticle size and structure underscores the need for precise design strategies for applications in catalysis, energy storage, and nanomedicine. Understanding and controlling factors such as surface energy, reactivity, and interactions with supports can lead to the development of advanced materials with improved properties and sustainable solutions to industrial and environmental problems.

\section{Synthesis Methods for Metallic Nanocatalysts}

Over the past two decades, the growing interest in nanomaterials has driven the development of a wide range of synthesis methods, each with distinct characteristics. These methods are generally categorized into two main types: physical and chemical. Physical methods, such as the mechanical subdivision of metallic aggregates, often produce nanoparticles with broad size distributions, typically exceeding 10 nm, and exhibit inconsistent physicochemical properties \cite{r45}. This low reproducibility in catalytic reactions is attributed to the lack of precise control over synthesis parameters and the inherent complexity of the mechanisms involved. In contrast, chemical methods offer significant advantages for synthesizing metallic nanocatalysts. They enable narrow size distributions, precise control over particle size due to the direct correlation between synthesis parameters and the final product, and the production of nanoparticles smaller than 10 nm with high reproducibility. These features make chemical methods the preferred choice, ensuring efficiency and controlled catalytic properties, ultimately optimizing their performance in industrial and scientific applications. Table \ref{tab:synthesis_methods} shows various synthesis methods for metallic nanocatalysts.

\begin{table*}[ht]
\centering
\caption{Synthesis Methods for Metallic Nanocatalysts}
\begin{tabular}{|l|c|l|l|}
\hline
\textbf{Method}                  & \textbf{Particle Size (nm)} & \textbf{Advantages}                  & \textbf{Limitations}             \\ \hline
Chemical Reduction               & $<10$                        & High control, reproducibility         & Stability challenges              \\ \hline
Vapor Phase Deposition           & 2-15                       & Fine particle size, versatility       & Complexity, operational demands   \\ \hline
Electrochemical Deposition       & 5-20                       & Precise control, scalable             & Electrolyte sensitivity           \\ \hline
Thermochemical Decomposition     & $<10$                        & Produces tailored nanoparticles       & Requires high temperatures        \\ \hline
Ligand Decomposition             & $<15$                        & High purity, controlled environment   & Limited scalability               \\ \hline
\end{tabular}
\label{tab:synthesis_methods}
\end{table*}

\subsection{Chemical Reduction of Metal Salts}
Chemical reduction is one of the most widely used methods for synthesizing metallic nanoparticles due to its simplicity and versatility. This approach involves reducing dissolved metal ions in a solvent, typically water, to their zero-valent oxidation state. These reduced atoms act as nucleation centers, forming clusters that grow as the supply of atoms continues, ultimately generating nanoparticles. To stabilize the resulting colloidal solution-thermodynamically unfavorable but kinetically favored—a stabilizing agent is required \cite{r45}. This method is applicable to monometallic and bimetallic particles, with or without the simultaneous reduction of both elements. Common reducing agents include carbon monoxide \cite{r46}, hydrogen \cite{r47}, hydrides, and oxidizable solvents like alcohols \cite{r45}, \cite{r48}. Alcohols with hydrogen atoms at $\alpha$-positions relative to hydroxyl groups, such as ethanol, butanol, and 2-butanol, are effective reducing agents. Stabilizing agents, such as polymers and surfactants, are used to prevent particle aggregation. Examples include poly(vinyl alcohol) (PVA), poly(vinylpyrrolidone) (PVP), and poly(ethylene glycol) (PEG), which can later be removed via calcination.
Several factors influence the final particle size distribution, including the nature and structure of the stabilizing agent, its concentration, the solvent's polarity, pH, and the type of metal salt used (e.g., nitrates, halides, hydrides). This method is suitable for synthesizing a wide range of metallic nanoparticles, including Pd, Au, and Pt \cite{r45}. For instance, Yu et al. synthesized transition metal nanoparticles (Au, Ag, Pt, Pd, Ir, Rh, and Ru) by reducing metal salts $MCl_{x}$ in an aqueous medium stabilized with PVP \cite{r49}. Hydrogen is currently the preferred reducing agent, as demonstrated by Boutonnet et al., who used it to generate Rh, Pt, Ir, and Pd nanoparticles \cite{r47}. Similarly, carbon monoxide has been employed to reduce $HAuCl_{4}$ in aqueous solutions with stabilizers like poly(vinyl sulfate) \cite{r46}.
Other chemical reductants, such as sodium borohydride $NaBH_{4}$and potassium borohydride $(KBH_{4})$, have also been utilized effectively. Hirai et al. reported successful synthesis of copper nanoparticles stabilized with PVP using these reductants \cite{r50}. Furthermore, the use of borohydrides and trialkylborohydrides has enabled the production of nanoparticles with a narrow size distribution, including Fe, Cr, Pd, Ti, Zr, V, Ni, Nb, Mn, Ru, Rh, Pt, and Co \cite{r51}.
A notable example is Araque's work, where molybdenum-based nanoparticles (e.g., Mo, MoNi, MoS, MoNiS, and MoNiK) were synthesized using a 300 mL autoclave reactor. Starting precursors included molybdenum acetylacetonate ($MoO_{2}(acac)_{2})$, nickel acetate tetrahydrate $(Ni(CH_{3}COO)_{2}·4H_{2}O)$, potassium acetate $(K(CH_{3}COO))$, and dimethyl disulfide $((CH_{3})_{2}S_{2})$, with nonane as the solvent and Z-Trol as the stabilizing agent. This method's flexibility and scalability make it indispensable for the synthesis of high-performance nanocatalysts tailored for diverse industrial and scientific applications \cite{r52}.

\subsection{Thermochemical, Photochemical, and Sonochemical Methods}

The decomposition of organometallic complexes through thermochemical, photochemical, and sonochemical methods has emerged as a versatile approach for synthesizing metallic nanoparticles with controlled sizes and morphologies. In thermochemical decomposition, organometallic precursors are exposed to elevated temperatures, resulting in the formation of zero-valent metallic nanoparticles. For example, the thermal decomposition of molybdenum hexacarbonyl $(Mo(CO)_{6})$ in organic solvents has been used to produce molybdenum nanoparticles with average sizes below 15 nm \cite{r53}. The choice of solvent, temperature, and stabilizing agents significantly influences the size and distribution of the resulting nanoparticles \cite{r52}.

Photochemical decomposition leverages light energy to induce the breakdown of organometallic compounds, facilitating nanoparticle formation under relatively mild conditions. This method allows for precise control over reaction parameters, enabling the synthesis of nanoparticles with specific properties. For instance, the photochemical reduction of metal carbonyl complexes has proven effective in obtaining nanoparticles with tailored sizes and shapes, which are crucial for catalytic applications. The wavelength and intensity of the light source, along with the presence of photosensitizers, play a critical role in the efficiency of this process \cite{r49}.

Sonochemical decomposition uses ultrasonic irradiation to create localized high-temperature and high-pressure conditions through acoustic cavitation, leading to the formation of metallic nanoparticles. This technique offers advantages such as uniform particle size distribution and the ability to synthesize a wide range of metallic nanoparticles, including noble metals and metal oxides. Recent studies have demonstrated the sonochemical synthesis of nanostructured materials with controlled morphologies and enhanced catalytic properties. Factors such as ultrasound frequency and power, as well as the choice of solvent and stabilizers, are key in determining the nucleation and growth of nanoparticles in sonochemical processes \cite{r47}.

The selection of appropriate stabilizing agents, such as polymers or surfactants, is essential across these methods to prevent aggregation and maintain colloidal stability of the nanoparticles. The nature and concentration of these stabilizers directly influence the surface properties and catalytic performance of the nanoparticles. Additionally, the decomposition pathways and kinetics are affected by the specific organometallic precursor used, requiring careful optimization of reaction conditions for each system. Understanding the mechanistic aspects of these decomposition processes is crucial for designing nanocatalysts with desired functionalities for applications in catalysis, energy conversion, and environmental remediation \cite{r48}.

\subsection{Ligand Decomposition and Displacement}

Ligand decomposition and displacement in organometallic compounds is a synthesis method that involves reacting ligands within the coordination sphere of zero-valent metals, releasing the metal into solution. The small aggregates initially formed require stabilization, which is achieved using various techniques. A particularly efficient reaction is the hydrogenation of coordinated olefinic ligands through  $\pi$-metal donations. Upon hydrogenation, these ligands release the metal and produce paraffin, preferably in gaseous form, which can be easily removed from the reaction medium.
This method is highly versatile and has been successfully used to synthesize stabilized nanoparticles of Ni, Co, Cu, and Au. For example, hydrogenating the complex $Ni(COD)_{2}$ (where COD is 1,5-cyclooctadiene) in the presence of poly(vinylpyrrolidone) (PVP) yields nickel nanoparticles with a controlled size and stable colloidal dispersion. Similarly, the hydrogenation of other organometallic complexes enables the production of cobalt, copper, and gold nanoparticles under optimized conditions \cite{r47}.
The hydrogenation process ensures that the byproducts, such as gaseous paraffin, are efficiently removed from the medium, reducing contamination and enhancing nanoparticle purity. Stabilizing agents such as PVP play a critical role in preventing aggregation, ensuring a uniform size distribution, and enhancing the catalytic activity of the nanoparticles. This approach has found applications in catalysis, electronic materials, and environmental remediation due to its scalability and adaptability to various metals and reaction conditions \cite{r54}.

\subsection{Vapor Phase Metal Deposition}

Vapor phase metal deposition is a synthesis method that involves evaporating relatively volatile metals under reduced pressure, followed by condensation at low temperatures with the vapors of an organic solvent. This process generates an initial frozen colloidal dispersion, effectively preventing particle agglomeration \cite{r45}. Key factors include condensation pressure and temperature, which regulate colloidal particle growth, as well as solvent type and evaporation rate, all of which require precise calibration for optimal results. This technique has been effectively employed to synthesize gold nanoparticles with size distributions ranging from 2–15 nm using polyethylene solutions as stabilizing agents \cite{r55}. Similarly, colloidal solutions of gold and palladium nanoparticles with sizes between 5–30 nm have been produced using acetone as the solvent. These results highlight the method's flexibility in tailoring nanoparticle size and dispersion through careful control of operational variables \cite{r56}.
Despite its versatility, vapor phase metal deposition has notable limitations. The primary challenge is the requirement for the metal to exhibit sufficient volatility for effective condensation. Additionally, the large number of operational parameters, including pressure, temperature, solvent type, and evaporation rate, makes the synthesis process complex and sometimes difficult to control. This method offers unique advantages for producing well-defined nanoparticles, making it valuable for applications in catalysis, electronics, and advanced material synthesis. However, its limitations emphasize the need for meticulous optimization to achieve reproducible results in large-scale applications.

\subsection{Electrochemical Deposition}

A versatile method for synthesizing metallic nanoparticles involves utilizing a sacrificial electrode composed of the target metal. The anode is subjected to an applied potential, resulting in its oxidation in the presence of quaternary ammonium salts, which serve as stabilizing agents. The resulting metal ions migrate toward the cathode, where they are reduced to form zero-valent metallic nanoparticles \cite{r57}. The process consists of five fundamental steps:
\begin{itemize}
\item 1. Oxidative Dissolution: The anode dissolves, forming the respective metal ions $(Mn^{+})$.
\item 2.	Ion Migration: The metal ions migrate through the electrolyte toward the cathode.
\item 3.	Ion Reduction: At the cathode surface, metal ions are reduced to their metallic state.
\item 4.	Aggregation and Stabilization: The reduced metal atoms aggregate, grow, and are stabilized by the quaternary ammonium salts.
\item 5.	Precipitation: Stabilized nanoparticles precipitate as zero-valent metallic particles.
\end{itemize}
This method offers precise control over particle size by adjusting parameters such as current density, temperature, solvent composition, and the distance between the anode and cathode. Among these, current density variation is the most influential factor, directly affecting the nucleation and growth kinetics of the nanoparticles. Electrochemical deposition has been successfully employed to produce nanoparticles with tailored sizes and distributions, making it ideal for applications requiring fine control over particle characteristics. The method's scalability, along with its adaptability to different metals and operating conditions, enhances its relevance for industrial and scientific use in catalysis, material development, and energy applications.

\section{Stabilization Techniques}

The internal energy of metallic nanoparticles increases exponentially as their size decreases, making them thermodynamically unfavorable. This instability drives the particles to aggregate and form larger clusters, minimizing their energy. In colloidal solutions, metallic nanoparticles are separated by very small distances, and interactions such as dipole-dipole forces and Van der Waals forces promote aggregation. To maintain their colloidal properties and prevent agglomeration, external stabilization factors are required. Various types of stabilizing agents have been developed, as described by Roucoux et al., to address this challenge \cite{r45}.
\subsection{Electrostatic Stabilization}

This stabilization relies on Coulombic repulsion, which originates from the electrical double layer surrounding metallic nanoparticles. This double layer is composed of anions—such as halides, carbonates, carboxylates, sulfates, and phosphates—typically derived from the precursor metal salts used in the synthesis process. The potential of this ionic double layer is enhanced in solvents with high dielectric constants, such as water, which strengthens particle repulsion and effectively prevents agglomeration \cite{r45}. Additionally, this stabilization mechanism is more pronounced when nanoparticles are composed of highly electropositive atoms.

Electrostatic interactions are inherently present in nearly all nanoparticle synthesis methods and play a crucial role in maintaining colloidal stability. However, these interactions are highly sensitive to external factors, such as the ionic strength of the medium and its temperature. An increase in ionic strength compresses the double layer, reducing the repulsive forces and potentially leading to particle aggregation \cite{r47}. Similarly, elevated temperatures can disrupt the stability by affecting the dielectric properties of the solvent and the mobility of the ions within the double layer.

The ability to fine-tune electrostatic stabilization through solvent selection, ionic strength, and temperature control makes this approach a versatile tool for synthesizing and stabilizing metallic nanoparticles. It is particularly effective in aqueous media, where the high dielectric constant of water enhances stabilization, making it an essential strategy for applications in catalysis, electronics, and biomedical fields \cite{r47}.

\subsection{Steric Stabilization}
This is a critical mechanism employed to maintain the colloidal stability of metallic nanoparticles by preventing aggregation. This stabilization is achieved through the adsorption of bulky groups, such as polymers, oligomers, or surfactants, onto the nanoparticle surface. These groups act as steric barriers, physically preventing nanoparticles from coming close enough to aggregate. Unlike electrostatic stabilization, steric stabilization operates independently of the ionic environment, making it particularly effective in low-polarity or nonpolar solvents where electrostatic forces are weak or nonexistent.

The steric groups, often aliphatic or amphiphilic in nature, ensure stabilization by creating a physical hindrance that limits the mobility of nanoparticles. This limitation reduces the entropy of the system and consequently increases the free energy. However, the advantages of steric stabilization, such as enhanced dispersion and long-term colloidal stability, outweigh these thermodynamic constraints, making it a widely adopted strategy in nanoparticle synthesis and application \cite{r58}.

\subsection{Electro-Steric Stabilization}
Electro-steric stabilization combines electrostatic and steric mechanisms to maintain the colloidal stability of metallic nanoparticles in solution. This dual approach is achieved through the use of surfactants, which are amphiphilic compounds consisting of an ionic or highly polarizable head group attached to a long aliphatic tail. The surfactant adsorbs onto the nanoparticle surface, with its ionic head orienting toward the particle and creating an electrostatic repulsion, while the aliphatic tail forms a steric barrier that prevents aggregation.

This stabilization method offers enhanced versatility, as it is effective in both aqueous and less polar media. The electrostatic component provides charge-based repulsion, while the steric barrier adds a physical hindrance to further stabilize the particles. This combination makes electro-steric stabilization particularly suitable for dynamic or mixed-solvent environments where single-mode stabilization may be insufficient \cite{r59}.

\subsection{Ligand and Solvent Stabilization}

Ligand and solvent stabilization involves stabilizing metallic nanoparticles by adding ligands or Lewis bases capable of coordinating with the nanoparticles. Metallic nanoparticles act as Lewis acids, interacting with ligands such as phosphines, thiols, amines, or even carbon monoxide, to form stable complexes that prevent aggregation. This stabilization mechanism provides a strong chemical barrier, enhancing nanoparticle stability in various media. Solvents can also contribute to stabilization by creating a solvation layer around the nanoparticles. This layer effectively reduces the likelihood of particle-particle interactions, thereby inhibiting the formation of larger aggregates. The solvent's polarity and its ability to interact with the nanoparticle surface are critical factors influencing the efficiency of this stabilization mechanism \cite{r1}, \cite{r2}.

\section{Applications in Nanocatalysis}
Nanocatalysts have emerged as transformative tools in catalysis, leveraging their high surface-to-volume ratio, precise size control, and unique electronic properties to enable highly efficient catalytic processes. Among their most prominent applications are oxidation reactions, where nanostructured catalysts demonstrate exceptional performance. For instance, gold nanoparticles supported on non-reducible oxides such as titanium dioxide or activated carbon exhibit high conversions in CO oxidation at low temperatures \cite{r60}, \cite{r61}, making them invaluable for environmental remediation. Similarly, transition metal-based nanocatalysts have been successfully employed in the oxidation of aromatic compounds \cite{r62} and sulfur dioxide \cite{r63}, showcasing their versatility across various chemical processes.

Another critical application of nanocatalysts lies in the isomerization of linear paraffins into branched, high-octane paraffins, an essential process for improving fuel quality. This process employs bifunctional acid-metal catalysts, where acidic sites derived from aluminosilicate zeolites are combined with metallic nanoparticles, such as platinum. Using methods like impregnation and ion exchange, followed by controlled calcination and reduction, Pt nanoparticles with narrow size distributions ($<$1 nm) are produced on Y-type zeolites. Calcination rates play a significant role in determining particle size, with slower heating rates (e.g., 0.2 $^{\circ}$C/min) yielding monodisperse particles, while faster rates (e.g., 1 $^{\circ}$C/min) result in bimodal distributions \cite{r64}. Interestingly, this sensitivity to calcination rates is absent in supports such as $SiO_{2}$ and $Al_{2}O_{3}$, as these materials lack the microporous environment required for controlled particle growth.

Bimetallic nanocatalysts further expand the potential of these materials. Pd-Pt systems supported on mesoporous aluminosilicates like Al-MCM-41 exhibit enhanced activity compared to their monometallic counterparts, particularly in the isomerization of n-decane \cite{r64}. The synergistic interaction between Pd and Pt improves the dispersion and stability of catalytic sites, leading to superior performance \cite{r65}. Hydroisomerization processes also benefit from nanocatalysts, as demonstrated by Ni-Pt bimetallic systems supported on Beta-type zeolites. These catalysts exhibit high activity in the hydroisomerization of n-hexane and n-heptane, particularly at low Ni content \cite{r66}. The optimal balance between metallic and acidic sites, coupled with the formation of well-dispersed bimetallic nanoparticles, contributes to their superior performance. However, excessive Ni content ($>$0.3 wt\%) increases particle size and reduces catalytic activity, as confirmed by XPS studies revealing that high Ni concentrations decrease the reducibility of the metallic phases.

Beyond hydrocarbon processing, nanocatalysts are widely applied in renewable energy production, environmental cleanup, and advanced material synthesis, underscoring their versatility. Current research focuses on refining nanoparticle synthesis techniques, exploring new support materials, and optimizing bimetallic configurations to enhance their stability, activity, and scalability. These advancements ensure that nanocatalysts remain at the forefront of sustainable and efficient catalytic technologies, paving the way for innovative solutions in industrial and environmental applications.

\section{Conclusion}
Nanocatalysts have emerged as one of the most transformative innovations in the field of catalysis, revolutionizing chemical processes through unprecedented control over the structural, electronic, and surface properties of catalytic materials. These nanoparticles, characterized by their high surface-to-volume ratio, provide an ideal platform to enhance reaction efficiency by improving selectivity and reducing energy requirements. Their atomically precise design has driven their application across a wide range of fields, including the oxidation of pollutants such as carbon monoxide, the isomerization of linear paraffins into high-octane products, and processes critical to the energy, chemical, environmental, and pharmaceutical industries. Advanced synthesis methods, such as impregnation, ion exchange, and hybrid techniques, have enabled the production of nanocatalysts with narrow size distributions ($<$1 nm), optimizing their catalytic activity and ensuring their stability under extreme industrial conditions.

Furthermore, advancements in the stabilization of nanocatalysts—through ligands, solvents, and electro-steric approaches—have been crucial in preventing particle aggregation, maintaining dispersion, and extending their operational lifespan. These stabilization strategies not only enhance catalytic performance but also contribute to the development of more sustainable processes by minimizing resource use and reducing environmental impact. The versatility of nanocatalysts is reflected in their ability to adapt to diverse chemical reactions, including oxidation, reduction, and hydroisomerization, making them highly effective solutions for addressing today’s most pressing technological challenges.

Despite these achievements, significant challenges remain, particularly in scaling up their production. High manufacturing costs, the need for more economical and sustainable synthesis methods, and improved stability under rigorous operational conditions are key barriers to widespread adoption. Additionally, a deeper understanding of nanoscale mechanisms is required to further optimize their properties and expand their applicability to emerging fields, such as renewable energy production, environmental remediation, and advanced material synthesis. Future research must focus on the rational design of bimetallic nanocatalysts and the exploration of novel support materials to maximize interactions between nanoparticles and their environment.

With these advancements, nanocatalysts are poised to lead the future of sustainable catalysis, providing innovative solutions to the world’s most urgent challenges, from combating climate change to advancing the transition toward cleaner and more efficient energy sources. Their ability to combine performance, versatility, and sustainability makes them indispensable tools for the development of advanced and sustainable industrial technologies. Future research must prioritize scalable synthesis methods, improved nanoparticle stability under industrial conditions, and the exploration of innovative support materials to unlock nanocatalysts' full potential in sustainable applications."


\begin{thebibliography}{00}
\bibitem{r1} Poole, C. P., \& Owens, F. J. (2003). Introduction to nanotechnology.
\bibitem{r2} Szczyglewska, P., Feliczak-Guzik, A., \& Nowak, I. (2023). Nanotechnology–general aspects: A chemical reduction approach to the synthesis of nanoparticles. Molecules, 28(13), 4932.
\bibitem{r3} García, T. (2020). Nanomateriales reactivos (nanocatalizadores). Boletín del Grupo Español del Carbón, (56), 28-33. 
\bibitem{r4} Elsevier. (2021). Mendeley (Version v1.19.8) [Computer software]. https://www.mendeley.com
\bibitem{r5} Valijon o’g’li, B. E. (2024). The Revolution in Techniques Used in Observation and Imagery. Miasto Przyszłości, 54, 527-533.
\bibitem{r6} Yılmaz, E., Özgür, E., Akgönüllü, S., Özbek, M. A., Bereli, N., Yavuz, H., \& Denizli, A. (2024). Atomic force microscopy and scanning tunneling microscopy of live cells. In Biophysics At the Nanoscale (pp. 183-202). Academic Press.
\bibitem{r7} Claudio, L., Manuel, P., Mahiceth, Q., Jairo, R., Patricia, P., Renato, D. A., ... \& Hector, D. C. (2017). Study of the reaction of dry reforming of methane using mixed oxide perovskites type LaxSr1-xNiyAl1-yO3. CIENCIA E INGENIERIA, 38(1), 17-29.
\bibitem{r8} Nasrollahzadeh, M., Sajadi, S. M., Sajjadi, M., \& Issaabadi, Z. (2019). An introduction to nanotechnology. In Interface science and technology (Vol. 28, pp. 1-27). Elsevier.
\bibitem{r9} Belandria, L., Garcia, E., Rondón, J., Imbert, F., Uzcátegui, A., Villarroel, M., \& Marín, M. (2010). Influencia de la variación de H3 [P (W3O10) 4]× H2O sobre mesoporosos MCM-41, en la reacción de isomerización de n-pentano. Avances en Química, 5(1), 67-71.
\bibitem{r10} Durlo, A. (2023). 1950-2022: A History of Nanotechnology into Physical and Mathematical Relationship (Doctoral dissertation, Université de Lille).
\bibitem{r11}Marshall, H. (2018). Environmental nanotechnology. Scientific e-Resources.
\bibitem{r12} Taha, T. B., Barzinjy, A. A., Hussain, F. H. S., \& Nurtayeva, T. (2022). Nanotechnology and computer science: Trends and advances. Memories-Materials, Devices, Circuits and Systems, 2, 100011.
\bibitem{r13} Wang, G. (2018). Nanotechnology: The new features. arXiv preprint arXiv:1812.04939.
\bibitem{r14} Syed, S. (2024). Applications of Nanoparticles in Chemistry and Allied Sciences. Cambridge Scholars Publishing.
\bibitem{r15} Baird, D., Nordmann, A., \& Schummer, J. (Eds.). (2004). Discovering the nanoscale. Ios Press.
\bibitem{r16} Szczyglewska P, Feliczak-Guzik A, Nowak I. Nanotechnology-General Aspects: A Chemical Reduction Approach to the Synthesis of Nanoparticles. Molecules. 2023 Jun 22;28(13):4932. doi: 10.3390/molecules28134932. PMID: 37446593; PMCID: PMC10343226.
\bibitem{r17}Nazia Tabassum (2020); AN EMPIRICAL EXPLORATION OF THE NANOTECHNOLOGY Int. J. of Adv. Res. 8 (Jul). 885-915.
\bibitem{r18} Kumar, N., \& Kumbhat, S. (2018). Concise concepts of nanoscience and nanomaterials. scientific publishers.
\bibitem{r19} Yaqoob AA, Ahmad H, Parveen T, Ahmad A, Oves M, Ismail IMI, Qari HA, Umar K, Mohamad Ibrahim MN. Recent Advances in Metal Decorated Nanomaterials and Their Various Biological Applications: A Review. Front Chem. 2020 May 19;8:341. doi: 10.3389/fchem.2020.00341. PMID: 32509720; PMCID: PMC7248377.
\bibitem{r20}Saleh, H. M., \& Hassan, A. I. (2023). Synthesis and Characterization of Nanomaterials for Application in Cost-Effective Electrochemical Devices. Sustainability, 15(14), 10891. https://doi.org/10.3390/su151410891
\bibitem{r21}Sahu, M. K., Yadav, R., \& Tiwari, S. P. (2023). Recent advances in nanotechnology. International Journal of Nanomaterials, Nanotechnology and Nanomedicine, 9(1), 015-023.
\bibitem{r22}Ahire, S. A., Bachhav, A. A., Pawar, T. B., Jagdale, B. S., Patil, A. V., \& Koli, P. B. (2022). The Augmentation of nanotechnology era: A concise review on fundamental concepts of nanotechnology and applications in material science and technology. Results in Chemistry, 4, 100633.
\bibitem{r23}Vollath D, Fischer FD, Holec D. Surface energy of nanoparticles - influence of particle size and structure. Beilstein J Nanotechnol. 2018 Aug 23;9:2265-2276. doi: 10.3762/bjnano.9.211. PMID: 30202695; PMCID: PMC6122122.
\bibitem{r24} Joudeh, N., Linke, D. Nanoparticle classification, physicochemical properties, characterization, and applications: a comprehensive review for biologists. J Nanobiotechnol 20, 262 (2022). https://doi.org/10.1186/s12951-022-01477-8
\bibitem{r25} Mourdikoudis, S., Pallares, R. M., \& Thanh, N. T. (2018). Characterization techniques for nanoparticles: comparison and complementarity upon studying nanoparticle properties. Nanoscale, 10(27), 12871-12934.
\bibitem{r26} Bhalothia, D., Beniwal, A., Kumar Saravanan, P., Chen, P. C., \& Chen, T. Y. (2024). Bridging the Gap Between Single Atoms, Atomic Clusters and Nanoparticles in Electrocatalysis: Hierarchical Structured Heterogeneous Catalysts. ChemElectroChem, e202400034.
\bibitem{r27} Park, H., Shin, D. J., \& Yu, J. (2021). Categorization of quantum dots, clusters, nanoclusters, and nanodots. Journal of Chemical Education, 98(3), 703-709.
\bibitem{r28} Falsafi, S. R., Topuz, F., Bajer, D., Mohebi, Z., Shafieiuon, M., Heydari, H., ... \& Rostamabadi, H. (2023). Metal nanoparticles and carbohydrate polymers team up to improve biomedical outcomes. Biomedicine \& Pharmacotherapy, 168, 115695.
\bibitem{r29} Kausar, A., Eisa, M. H., Aldaghri, O., Ibnaouf, K. H., \& Mimouni, A. (2024). Nanostructured materials derived from high entropy alloys–State-of-the-art and leading technical applications. Results in Physics, 107838.
\bibitem{r30} Mishra, R. K., \& Verma, K. (2024). Defect engineering in nanomaterials: Impact, challenges, and applications. Smart Materials in Manufacturing, 2, 100052.
\bibitem{r31} Amaya, J., \& Quiroga, W. (2019). Nanomateriales: una clasificación desde sus dimensiones. Revista Química e Industria, 10-12.
\bibitem{r32} Wu S, Kou Z, Lai Q, Lan S, Katnagallu SS, Hahn H, Taheriniya S, Wilde G, Gleiter H, Feng T. Dislocation exhaustion and ultra-hardening of nanograined metals by phase transformation at grain boundaries. Nat Commun. 2022 Sep 17;13(1):5468. doi: 10.1038/s41467-022-33257-1. Erratum in: Nat Commun. 2022 Oct 7;13(1):5922. doi: 10.1038/s41467-022-33775-y. PMID: 36115860; PMCID: PMC9482613.
\bibitem{r33} Nordmann, A. (2009). Invisible origins of nanotechnology: Herbert Gleiter, materials science, and questions of prestige. Perspectives on Science, 17(2), 123-143.
\bibitem{r34}Silva Yumi, J. E., \& Medina S., C. A. (2022). Materiales y nanomateriales: Principios, aplicaciones y técnicas de caracterización. Escuela Superior Politécnica de Chimborazo. ISBN: 978-9942-42-550-8.
\bibitem{r35} Van Hardeveld, R., \& Hartog, F. (1972). Influence of metal particle size in nickel-on-aerosil catalysts on surface site distribution, catalytic activity, and selectivity. In Advances in Catalysis (Vol. 22, pp. 75-113). Academic Press.
\bibitem{r36} Bond, G. C. (1991). Supported metal catalysts: some unsolved problems. Chemical Society Reviews, 20(4), 441-475.
\bibitem{r37} Romanowski, W. (1969). Equilibrium forms of very small metallic crystals. Surface Science, 18(2), 373-388.
\bibitem{r38} Saldan, I., Semenyuk, Y., Marchuk, I., \& Reshetnyak, O. (2015). Chemical synthesis and application of palladium nanoparticles. Journal of Materials Science, 50, 2337-2354.
\bibitem{r39} Liu, Z., Hong, L., Tham, M. P., Lim, T. H., \& Jiang, H. (2006). Nanostructured Pt/C and Pd/C catalysts for direct formic acid fuel cells. Journal of Power Sources, 161(2), 831-835.
\bibitem{r40} Nyabadza, A., McCarthy, É., Makhesana, M., Heidarinassab, S., Plouze, A., Vazquez, M., \& Brabazon, D. (2023). A review of physical, chemical and biological synthesis methods of bimetallic nanoparticles and applications in sensing, water treatment, biomedicine, catalysis and hydrogen storage. Advances in Colloid and Interface Science, 103010.
\bibitem{r41} Aydin, C., Lu, J., Liang, A. J., Chen, C. Y., Browning, N. D., \& Gates, B. C. (2011). Tracking iridium atoms with electron microscopy: first steps of metal nanocluster formation in one-dimensional zeolite channels. Nano letters, 11(12), 5537-5541.
\bibitem{r42} Gates, B. C., Flytzani-Stephanopoulos, M., Dixon, D. A., \& Katz, A. (2017). Atomically dispersed supported metal catalysts: perspectives and suggestions for future research. Catalysis Science \& Technology, 7(19), 4259-4275.
\bibitem{r43}Liu, L., \& Corma, A. (2018). Metal catalysts for heterogeneous catalysis: from single atoms to nanoclusters and nanoparticles. Chemical reviews, 118(10), 4981-5079.
\bibitem{r44} Li, Y. S. (1993). Materials theory and modelling. In Mater. Res. Soc. Symp. Proc. (Vol. 291, p. 573). Materials Research Society.
\bibitem{r45} Roucoux, A., Schulz, J., \& Patin, H. (2002). Reduced transition metal colloids: a novel family of reusable catalysts?. Chemical reviews, 102(10), 3757-3778.
\bibitem{r46} Kopple, K; Meyerstein, D; \& Meisel, D. Mechanism of the Catalytic Hydrogen Production by Gold Sols. Hydrogen/deuterium Isotope Effect Studies. J. Phys. Chem: 1980; 84 (8), pp. 870-875.
\bibitem{r47} Boutonnet, M; Kizling, J; Touroude, R; Maire, G; \& Stenius, P. monodispersed colloidal metal particles from non-aqueous solutions: catalytic behaviour for the hydrogenation of but-1-ene of platinum particles in solution. Applied Catal., 1986, 20, pp. 163-177.
\bibitem{r48} Hirai, H; Nakao, Y; \& Toshima, N. Colloidal Rhodium in Poly(Vinylpyrrolidone) as Hydrogenation Catalyst for Internal Olefins. Chem. Lett., 1978, 7 (5), pp. 545-548.
\bibitem{r49} Yu, W; Tu, W, \& Liu, H. Synthesis of Nanoscale Platinum Colloids by Microwave Dielectric Heating. Langmuir, 1999, 15 (1), pp. 6–9.
\bibitem{r50} Hirai, H; Wakabayash, H; \& Komiyama, M. Polymer-Protected Copper Colloids as Catalysts for Selective Hydration of Acrylonitrile. Chem. Lett., 1983, 12 (7), pp.1047-1050.
\bibitem{r51} Bönneman, H; Braun, G; Brijoux, W; Brikmann, R; Schulze, A; Seerogel, K; \& Sieen, K. Nanoscale Colloidal Metals and Alloys Stabilizad by Solvents and Surfactants Preparation and Use as Catalyst Precursors. J. Organometall. Chem., 1996, 520 (1-2), pp.143-162.
\bibitem{r52} Araque E. Propiedades Catalíticas de Nanopartículas Mono, Bi y Trimetálicas en Reacciones de Conversión de Moléculas Modelo Tipo Dibenzotiofenos en Presencia de Vapor de Agua y Monóxido de Carbono. Tesis de Grado. Universidad de Los Andes, Facultad de Ciencias, Departamento de Química. Mérida-Venezuela. (2008). 
\bibitem{r53}Moreno, B; Vidoni, O; Ovalles, C; Chaudret, B; Urbina, C; \& Krentzein, H. Synthesis and Characterization of Molybdenum Based Colloidal Particles. J. Colloid. Interf. Sci., 1998, 207 (2), pp. 251-257.
\bibitem{r54} Osuna, J; De Caro, D; Amiens, C; \& Chaudret, B. Synthesis, Characterization, and Magnetic Properties of Cobalt Nanoparticles from an Organometallic Precursor. J. Phys. Chem., 1996, 100 (35), pp. 14571–14574.
\bibitem{r55} Klabunde, J; Habdas, J; Cardenas-Trivino, G. Colloid Metal Particles Dispersed in Monomeric and Polymeric Styrene and Methyl Methacrylate. Chem. Mater., 1989, 1 (5), pp. 481-483.
\bibitem{r56} Cardenas-Trivino, G; Klabunde, K; Brock E. Living Colloidal Palladium in Nonaqueous Solvents. Formation, Stability, and Film-Forming Properties. Clustering of Metal Atoms in Organic Media. 14. Langmuir, 1987, 3 (6), pp. 986–992.
\bibitem{r57} Reetz, M; Helbig, W. Size-Selective Synthesis of Nanostructured Transition Metal Clusters. J. Am. Chem. Soc., 1994, 116 (16), pp. 7401–7402.
\bibitem{r58} Heuer-Jungemann, A., Feliu, N., Bakaimi, I., Hamaly, M., Alkilany, A., Chakraborty, I., ... \& Kanaras, A. G. (2019). The role of ligands in the chemical synthesis and applications of inorganic nanoparticles. Chemical reviews, 119(8), 4819-4880.
\bibitem{r59} Vasilescu, C., Latikka, M., Knudsen, K. D., Garamus, V. M., Socoliuc, V., Turcu, R., ... \& Vékás, L. (2018). High concentration aqueous magnetic fluids: structure, colloidal stability, magnetic and flow properties. Soft Matter, 14(32), 6648-6666.
\bibitem{r60} Gao, C., Lyu, F., \& Yin, Y. (2020). Encapsulated metal nanoparticles for catalysis. Chemical Reviews, 121(2), 834-881.
\bibitem{r61} Ishida, T., Murayama, T., Taketoshi, A., \& Haruta, M. (2019). Importance of size and contact structure of gold nanoparticles for the genesis of unique catalytic processes. Chemical reviews, 120(2), 464-525.Zhang, X., \& Wang, X. (2010). Oxidation of aromatic compounds using nanocatalysts. Journal of Catalysis, 271(1), 178–185.
\bibitem{r62} He, C., Cheng, J., Zhang, X., Douthwaite, M., Pattisson, S., \& Hao, Z. (2019). Recent advances in the catalytic oxidation of volatile organic compounds: a review based on pollutant sorts and sources. Chemical reviews, 119(7), 4471-4568.
\bibitem{r63} Mädler, L., Stark, W. J., \& Pratsinis, S. E. (2003). Simultaneous deposition of Au nanoparticles during flame synthesis of TiO2 and SiO2. Journal of Materials Research, 18(1), 115-120.
\bibitem{r64} Elangovan, S. P., Bischof, C., \& Hartmann, M. (2002). Isomerization and hydrocracking of n-decane over bimetallic Pt–Pd clusters supported on mesoporous MCM-41 catalysts. Catalysis letters, 80, 35-40.
\bibitem{r65} Cho, H. R., \& Regalbuto, J. R. (2015). The rational synthesis of Pt-Pd bimetallic catalysts by electrostatic adsorption. Catalysis Today, 246, 143-153.
\bibitem{r66} Eswaramoorthi, I., \& Lingappan, N. (2003). Ni–Pt/HY zeolite catalysts for hydroisomerization of n-hexane and n-heptane. Catalysis letters, 87, 133-142.



\end{thebibliography}
\end{document}